\documentclass[a4paper,11pt]{article}
\usepackage{jcappub} 

\usepackage{amssymb,graphicx}

\newcommand{\beqn}{\begin{eqnarray}}
\newcommand{\eeqn}{\end{eqnarray}}
\newcommand{\beq}{\begin{equation}}
\newcommand{\eeq}{\end{equation}}

\def\simleq{\; \raise0.3ex\hbox{$<$\kern-0.75em
      \raise-1.1ex\hbox{$\sim$}}\; }
   
\def\simgeq{\; \raise0.3ex\hbox{$>$\kern-0.75em
      \raise-1.1ex\hbox{$\sim$}}\; }
      
\newcommand{\la}{\langle}

\newcommand{\ra}{\rangle}

\renewcommand{\(}{\left(}

\renewcommand{\)}{\right)}

\begin{document}

\title{Beginning inflation in an inhomogeneous universe}
\author[a]{William E.\ East,}
\affiliation[a]{Kavli Institute for Particle Astrophysics and Cosmology, Stanford
University, SLAC National Accelerator Laboratory, Menlo Park, California 94025,
USA}
\author[b]{Matthew Kleban,} 
\affiliation[b]{ Center for Cosmology and Particle Physics, New York University} 
\author[c]{Andrei Linde,} 
\affiliation[c]{ SITP and Department of Physics, Stanford University, Stanford,
California 94305, USA }
\author[c,a]{and Leonardo Senatore} 

\abstract{
Using numerical solutions of the full Einstein field equations coupled to a
scalar inflaton field in 3+1 dimensions, we study the conditions under which a
universe that is initially expanding, highly inhomogeneous and dominated by gradient
energy can transition to an inflationary period.  If the initial scalar field
variations are contained within a sufficiently flat region of the inflaton
potential, and the universe is spatially flat or open on average,  inflation will occur
following the dilution of the gradient and kinetic energy due to expansion.
This is the case even when the scale of the inhomogeneities is comparable to the
initial Hubble length, and overdense regions collapse and form black holes,
because underdense regions continue expanding, allowing inflation to eventually
begin.  This establishes that inflation can arise from highly
inhomogeneous initial conditions and solve the horizon and flatness problems, at
least as long as the variations in the scalar field do not include values that
exceed the inflationary plateau.
}

\maketitle

\section{Introduction}
Inflation was originally proposed to account for the high degree of flatness and
homogeneity observed today in our universe at horizon
scales~\cite{Guth:1980zm,Linde:1981mu,Albrecht:1982wi,Linde:1983gd}.  However,
an oft-repeated criticism is that, while it is true that \emph{if} inflation
starts, it rapidly inflates away any inhomogeneities, in order to begin
inflation requires homogeneity over several or many Hubble volumes, because
otherwise the expanding universe will be increasingly dominated by
inhomogeneities rather than inflationary potential energy.  Hence, how can
inflation explain homogeneity on Hubble scales today?

Here we address to what degree inflation requires initial homogeneity. Arguably
this problem does not appear if inflation can start at nearly Planckian
density~\cite{Linde:1983gd,Linde:1985ub}.  There are also several methods for
solving the problem of initial conditions for low energy scale inflation,
see~\cite{Zeldovich:1984vk,Cornish:1996st,Coule:1999wg,Linde:2004nz,Linde:2014nna,Carrasco:2015rva}
and references therein.  
There has also been work establishing ``no hair" type results governing the
decay of perturbations for spacetimes with a cosmological constant, including
important work showing how to construct a large class of solutions by imposing
the requirement that the solution approach de Sitter at late
times~\cite{1983ZhPmR..37...55S,fefferman1985conformal}. A review of this can
be found in~\cite{Anninos:2012qw}.
Here we concentrate on single-field inflation models
where the energy scale of inflation is orders of magnitude smaller than the
Planck energy, which have become more popular in the light of observational
data~\cite{Planck:2013jfk}.  Since small perturbations certainly do not prevent
inflation from beginning, the cases of interest are those with large initial
inhomogeneities, outside the regime of linear theory and requiring numerical
solutions.

This question has been studied in the past using general relativistic
simulations in 1D~\cite{Albrecht:1986pi,KurkiSuonio:1987pq}
and spherical symmetry~\cite{Goldwirth:1989pr} (see
also~\cite{KurkiSuonio:1993fg} for early work in 3D), as well as in the
multi-field inflation case but ignoring inhomogeneities in the gravity
sector~\cite{Easther:2014zga}.  
In this work, we solve the full, nonlinear
Einstein equations numerically including cases with large inhomogeneities that
give rise to collapsing regions and trapped surfaces.  We find that many
cosmologies that are initially expanding, highly inhomogeneous and strongly dominated by
spatial gradient energy give rise to inflation.  This happens because while
overdense regions collapse into black holes, the 
underdense regions
evolve into voids that become dominated by inflationary vacuum energy at (much)
later times.  

Our result fundamentally alters the interpretation of analyses such as~\cite{PhysRevD.61.023502}, 
which places a lower bound on the size of an inflationary patch when it eventually emerges, requiring
it to be larger than the background Hubble radius.
We demonstrate that such inflationary patches can arise
simply through gravitational collapse and redshifting from conditions
that, one or more Hubble times  prior to the onset of inflation, were very inhomogeneous.  
This is a dynamical mechanism that does not require any acausal interaction
between regions initially separated by more than one horizon distance.

This picture holds when the range of initial scalar field values lies entirely
on the ``inflationary plateau" where the potential is flat, satisfying
$\epsilon_{n} \equiv \left( M_{P} \partial_{\phi}\right)^{n }\ln V \ll 1$ (where
$M_P$ is the Planck mass), with the number of inflationary e-folds in
homogeneous universe scaling as $N \sim \Delta \phi / M_{P} \epsilon_{1}$
(assuming $\dot \phi$ is small initially, but a closely related statement can be
made  when $\dot \phi$ is large, see e.g.~\cite{Carrasco:2015rva}).  In the
particular class of so-called cosmological attractors (see
\cite{Linde:2014nna,Carrasco:2015rva} and references therein), which have
potentials that rapidly approach a positive constant for large values of the
inflaton field, this plateau includes all but a finite range of values -- and
hence, with high probability the average value of the field falls in a range
where the potential is almost exactly a cosmological constant.  Nevertheless, we
also consider the complimentary case where the spatially averaged value of the
scalar field $\langle \phi \rangle$ lies on the plateau, but the variations in
the field fall outside this plateau.  A na\"{i}ve expectation would be that
these would also undergo inflation.  Instead, we find that when the fluctuations
around the average exceed the distance in field space to the end of the plateau,
$\delta \phi > \Delta \phi$, inflation sometimes does not take place regardless
of the value of $\langle \phi \rangle$.  This is because the field feels the
effect of the potential beyond the plateau in at least some parts of the
universe, and --- depending on the potential --- this can have a tendency to
pull $\langle \phi \rangle$ strongly towards the  minimum, which is necessarily
off the plateau.  This condition is of course completely invisible in
near-homogeneous cosmologies where $\delta \phi$ is small or vanishing.

\begin{figure}
\begin{center}
\includegraphics[angle=90,width=0.75\columnwidth,draft=false]{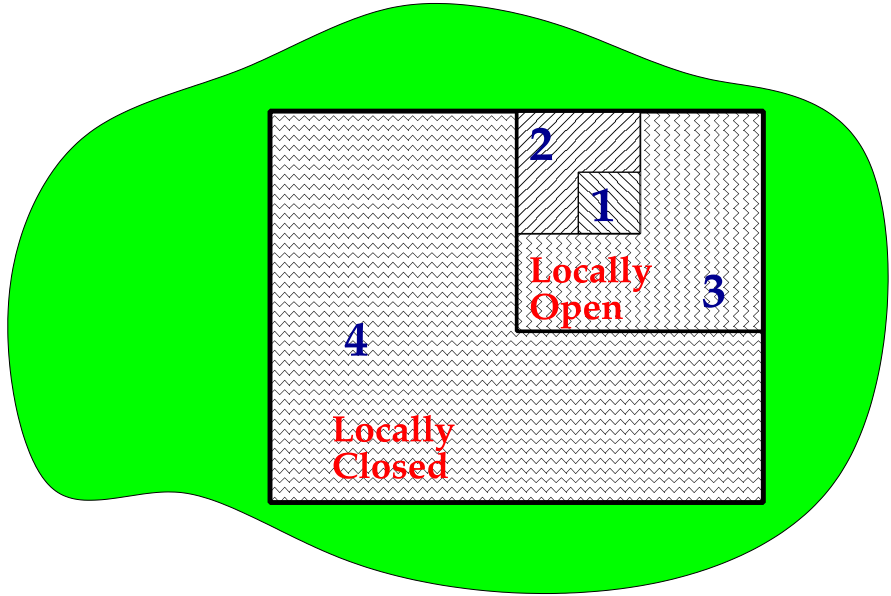}
\end{center}
\caption{
Let us consider a region (labelled ``(4)") of size comparable of the
would-be inflationary Hubble patch. If the region is open or flat on average, it
means that modes longer than the region will not make the whole region collapse. For
each subdivision of region (4), there has to be one region, here labelled (3),
where the role of longer modes is to make the region open or flat, while its
complement (the part of (4) not within (3)) can be closed.  The same applies to
further subdivisions, here (2) and (1). In this way, given a region (4) that is
flat or open on average, there is a sequence of smaller region where the effect
of long modes is to keep it open at all times.  Therefore, long modes will not
prevent geodesics originating in region (1) from eventually inflating.  As our
simulations show, modes shorter than the horizon at any given time can produce
localized black holes, but do not make the region as a whole collapse.
\label{fig:open-within-open}
}
\end{figure}

Our conclusions are based on a set of fully general relativistic simulations in
3+1 dimensions.  The field content is a single scalar $\phi$ with a potential
and average value that would allow for 60 e-folds or more of inflation given
homogeneous initial conditions.  However, we begin with inhomogeneous
conditions where the energy in spatial gradients dominates over the potential
energy by a factor of $10^3$.  We simulate an initially expanding universe with
toroidal topology where the nonzero wavenumbers $k$ of the initial scalar field
content satisfy $k/H_{0} \geq 1$ (i.e. have wavelength $\leq 2 \pi H_0^{-1}$),
where $H_{0}$ is the expansion rate on the initial time slice.  Without
simulating them, the behavior of longer wavelength modes can be understood as
follows. When $k/H \ll 1$ (a ``long" mode), the mode simply renormalizes the
local density; that is, so long as they remain outside the horizon, the effect
of long modes is captured by homogeneous cosmologies, which are well
understood~\cite{Wald:1983ky}.  In fact, the effect of long modes on a volume
that has been expanding at some arbitrary initial time will either be to make
it closed, in which case it collapses after a finite time, or open, in which
case the volume keeps expanding.  In a universe that is ``flat (or open) on
average,'' it is easy to see that there must exist at least one region where
the effect of long modes is to keep it open at all times.  This is illustrated
in Fig.~\ref{fig:open-within-open}.  Once a long mode re-enters the horizon its
evolution is much more complex, but is then described by our simulations. 

Instead of considering the periodicity of our simulations as a convenient
boundary condition, a complementary scenario covered by this setup is to imagine
that the universe is a torus of length of order $2\pi H_0^{-1}$, with
perturbations with wavenumber of order $H_0$, which could all be on the order of
the Planck
scale~\cite{Zeldovich:1984vk,Cornish:1996st,Coule:1999wg,Linde:2004nz,Linde:2014nna,Carrasco:2015rva}.
With this interpretation, our simulations describe the whole universe rather
than a limited part.

\section{Methodology}
In order to simulate an inhomogeneous cosmology, we solve the Einstein field
equations coupled to the inflaton $\phi$, which has equation of motion $\Box
\phi = V'$.  We evolve the field equations in the generalized harmonic
formulation using the code described in~\cite{Pretorius:2004jg, code_paper}.  We
use units where $8\pi G=c=1$ throughout. 

For the scalar field initial conditions, we choose a superposition of standing
waves with $\dot{\phi}(t=0)=0$ and 
\begin{equation} 
\phi(t=0,\mathbf{x}) =
\phi_0 + \delta \phi \left[ \sum_{1\leq |\mathbf{k}L/2 \pi|^2 \leq N }\hskip
-0.5cm  \cos(\mathbf{k}\cdot \mathbf{x}+\theta_{\mathbf{k}}) \right],
\label{eq:phi_ic} 
\end{equation}
where $\mathbf{k}$ ranges over all the allowed wavevectors in a
three-dimensional periodic domain (e.g. for $N=1$, the six wavevectors given by
plus and minus each of the three coordinate directions), and the
$\theta_{\mathbf{k}}$ values are randomly chosen phases. 
We always take the length of the domain in each direction to be equal to the
wavelength of the longest mode in that simulation $L=2\pi/k_{\rm min}$, and
consider various ratios of this to the initial Hubble scale.  
We construct initial data for the metric by solving the
constraint equations using the code described in~\cite{idsolve_paper}. We assume
(at the initial time only) that the spatial metric is conformally flat $\gamma_{ij}=\Psi^4 f_{ij}$, and
that the trace of the extrinsic curvature $K$ is constant while the traceless
part is zero, so that the momentum constraint is trivially satisfied.  
The value
of the conformal factor $\Psi$ then comes from solving the Hamiltonian
constraint. 

We choose $K$ based on the integral of the Hamiltonian constraint
over the periodic domain (ignoring the conformal factor)
\begin{equation}
K = -\left[\frac{1}{V}\int\left(3 V(\phi)+\frac{3}{2}\partial_i \phi \partial^i
     \phi\right)dV\right]^{1/2},
     \label{eqK}
\end{equation} 
which we find to give solutions with $\Psi \sim 1$.  The choice of a negative
$K$ will give us an initially expanding universe
with $H_0=-K/3$.  (A positive $K$, on the other hand, would give an initially contracting universe, which would likely result in a crunching-type solution.) In the special case of a homogeneous scalar field
configuration ($\delta \phi=0$) this choice of initial data will reduce to a
time slice of an FRW solution.

We consider several inflationary potentials, the simplest being a
cosmological constant $V(\phi)=\Lambda$.  The next is
\begin{equation}
V(\phi) = \Lambda(1+\exp(-\lambda \phi))^{-1}
\label{eq:step_potential}
\end{equation}
which is flat for $|\phi| \gg 1/\lambda$ and approximates a step function for
large  $\lambda$.  The last is the ``notch" potential
\begin{equation}
V(\phi) = \Lambda \tanh^2(\lambda \phi)\ 
\label{eq:notch_potential}
\end{equation}
describing a family of cosmological
attractors~\cite{Kallosh:2013hoa,Kallosh:2013yoa}.  These potentials cover a
large enough class to allow our conclusions to be quite generic: in the flat
regions they are well described by a cosmological constant $V(\phi)=\Lambda$
or~$0$, while the different ways in which the potentials tend to zero cover a
wide range of possibilities.

We will make use of several quantities defined in terms of the timelike unit
normal to slices of constant coordinate time $n^a$ for characterizing the
simulations below.  The vector $n^a$ can be considered the four velocity of a
fiducial observer whose worldline is orthogonal to the constant time
hypersurfaces.  We will use the energy density $\rho = n_a n_b T^{ab}$ where
$T^{ab}$ is the scalar field stress-energy tensor. We will also calculate the
local volume expansion $\theta := \nabla_a n^a = -K$. This will allow us to
define a fiducial local Hubble expansion rate $H:=\theta /3$ and corresponding
number of $e$-folds of expansion $\mathcal{N}$ found by integrating $n^a
\nabla_a
\mathcal{N} = \theta/3$ (see e.g.~\cite{Xue:2013bva}).  We denote the volume
average of a quantity by 
$ \langle X \rangle := \int X \sqrt{\gamma}d^3x / \int \sqrt{\gamma} d^3x  $
where $\gamma = \det\gamma_{ij}$.
See the appendix
for details on the numerical method and resolution study results demonstrating
convergence. 

\section{Results}
We perform a number of simulations to study the conditions under which a
gradient energy dominated universe, $\langle \rho \rangle = 10^3 \Lambda$, will
transition to exponential expansion.  In addition to this ratio of gradient to
potential energy, we will quantify the inhomogeneities in the different cases we
consider by referring to the ratio of the wavenumbers of the initial scalar
field content to the expansion rate on the initial time slice $k/H_0$. 

We first present results using a
cosmological constant, and following that present cases with non-trivial
potentials where some part of the scalar field range falls outside the flat
region of the potential.  

\begin{figure}
\begin{center}
\includegraphics[width=\columnwidth,draft=false]{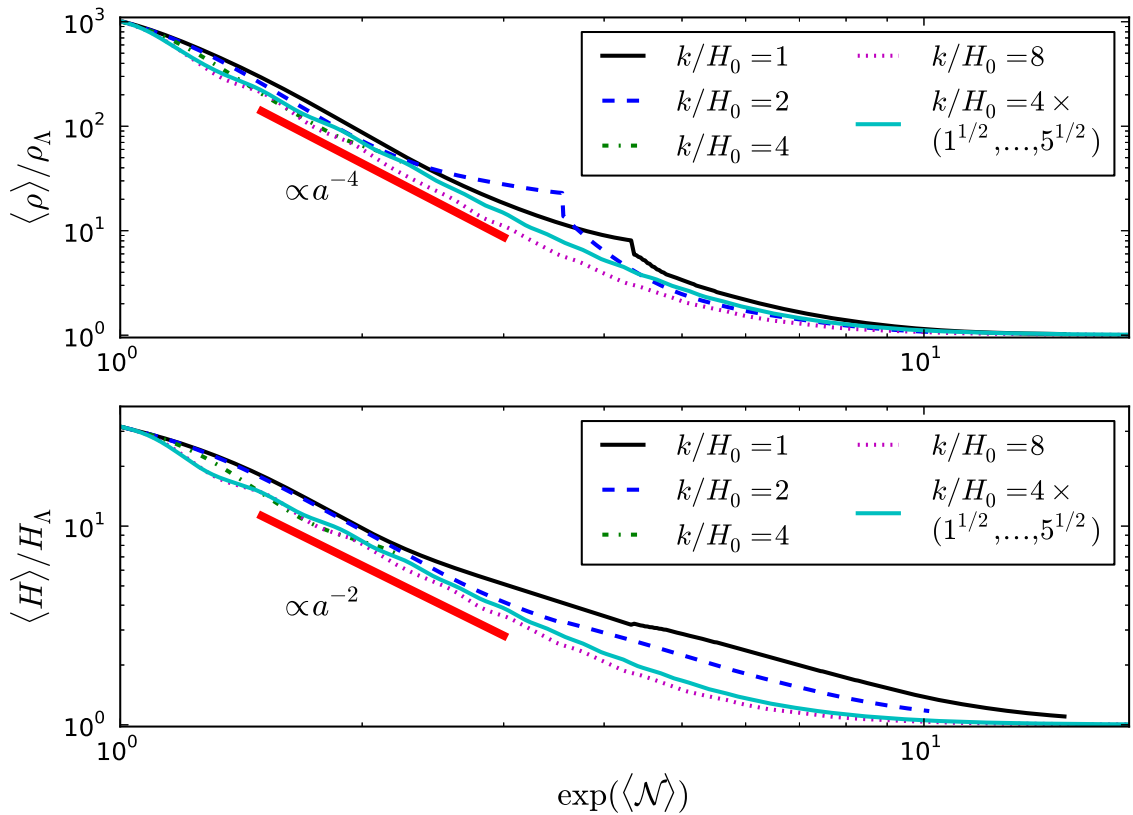}
\end{center}
\vspace{-10pt}
\caption{
The volume-averaged energy density (top) and expansion rate (bottom)
plotted against a volume-averaged measure of the effective scale factor
for cases with a cosmological constant potential. We show several
cases with inhomogeneities at a single wavenumber magnitude ($N=1$), as well
as one case with several different wavenumber magnitudes ($N=5$) where the
smallest wavenumber magnitude is $k/H_0=4$. For comparison, the slope of these
quantities expected in a radiation dominated FRW universe is also shown.  Once
apparent horizons are found, we ignore their interiors for the purpose of
calculating these quantities, which accounts for the discontinuous features
evident in the $k/H_0=1$ and 2 cases.
\label{fig:rhoH_a} }
\end{figure}

\begin{figure}
\begin{center}
\includegraphics[width=0.6\columnwidth,draft=false]{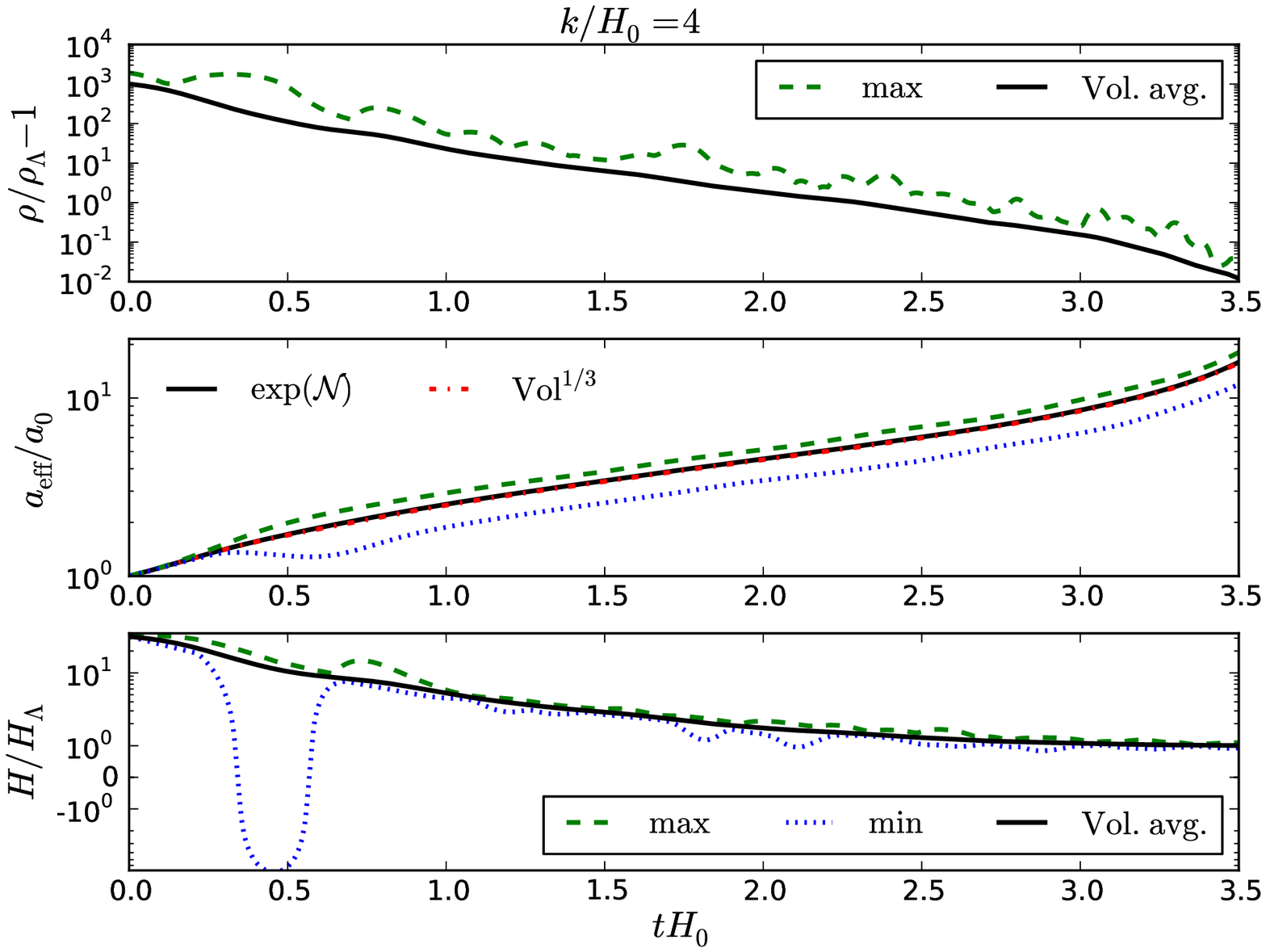}
\includegraphics[width=0.6\columnwidth,draft=false]{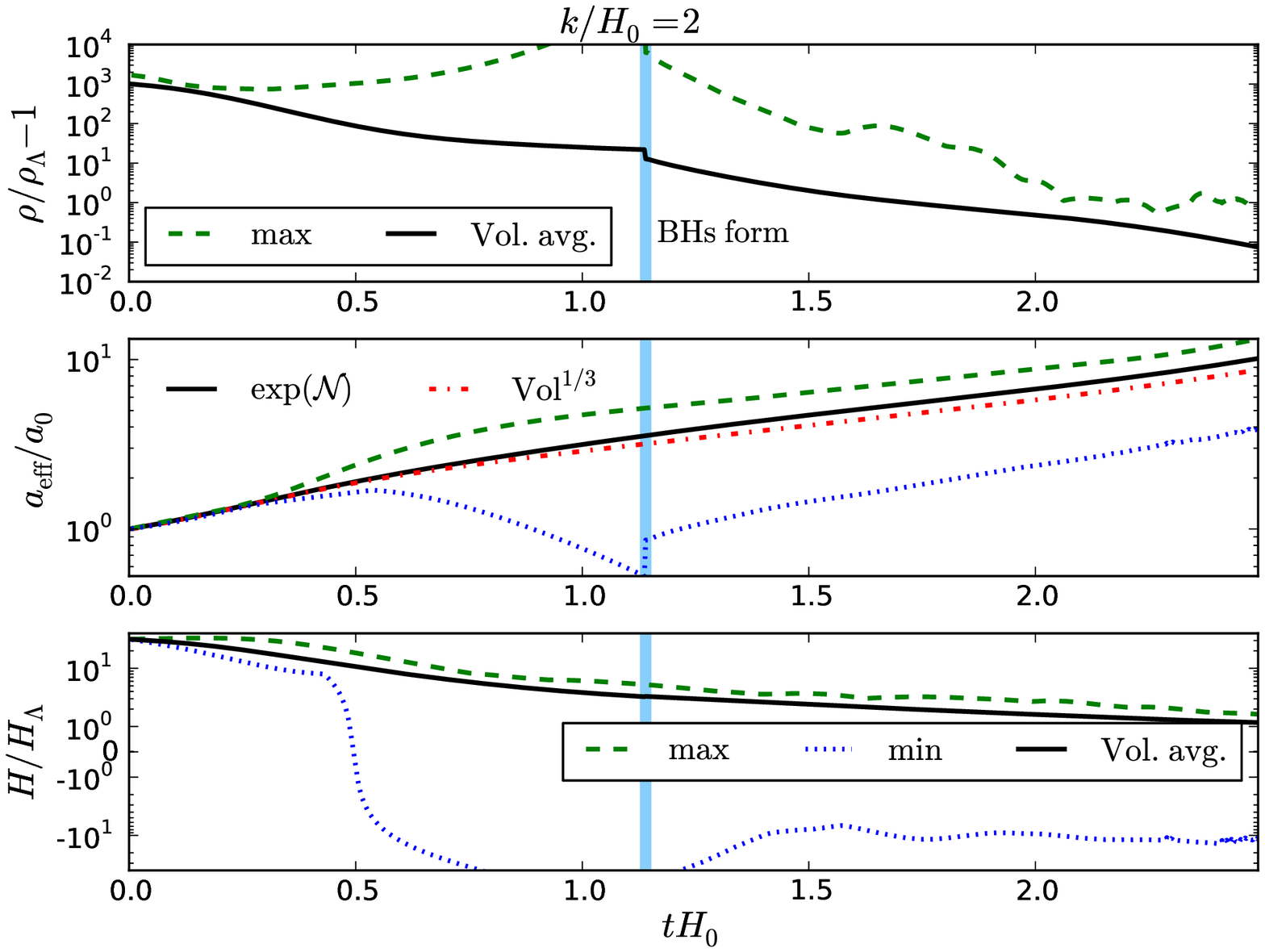}
\includegraphics[width=0.6\columnwidth,draft=false]{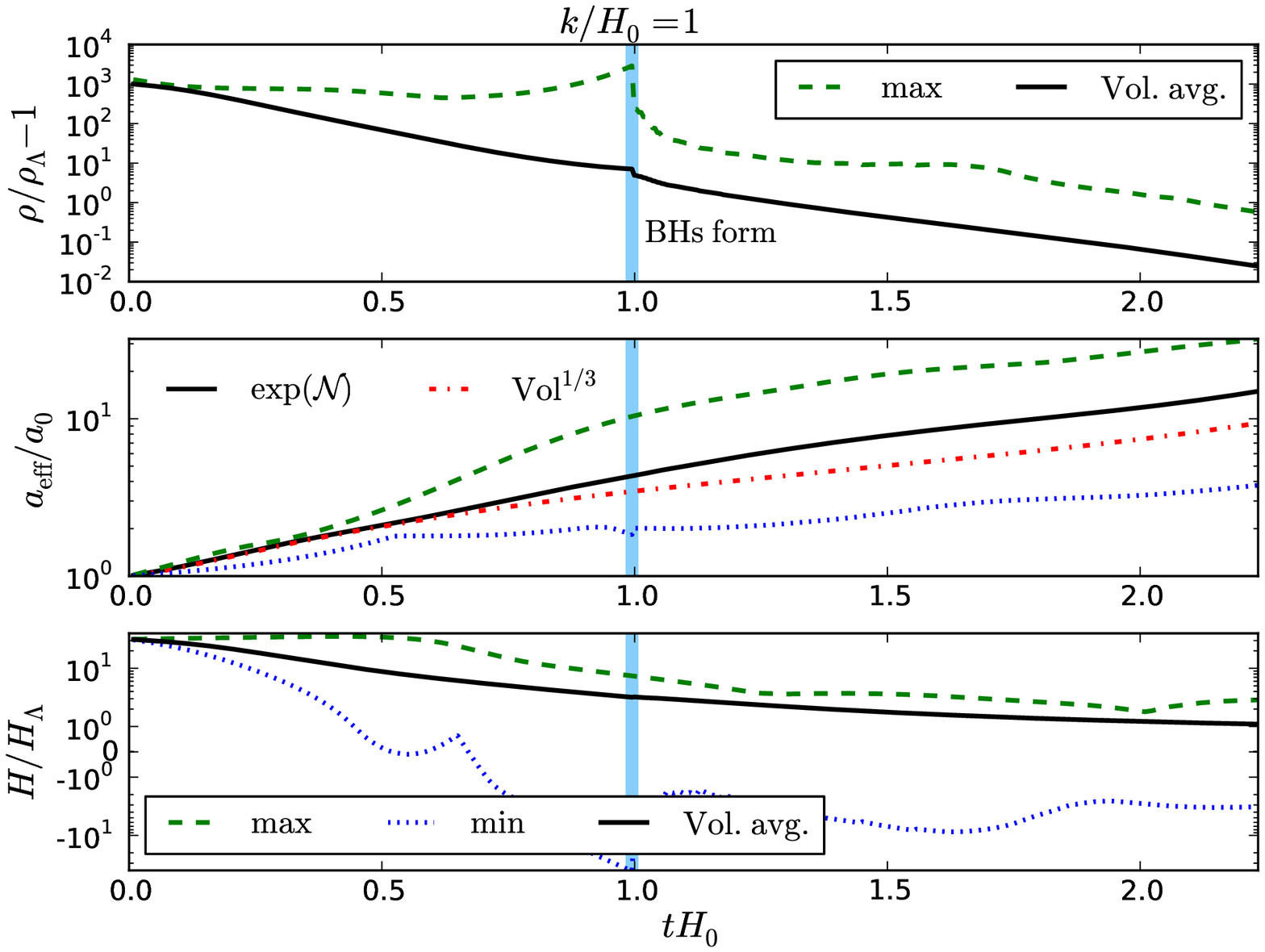}
\end{center}
\vspace{-10pt}
\caption{
Results from cases with a constant potential, $N=1$, and  
$k/H_0=4$, 2, and 1 (top to bottom). 
Each panel has three subplots which, from top to
bottom, show as a function of time: the energy density minus the vacuum energy
contribution, normalized by the vacuum energy density; a measure of the
effective scale factor (normalized to be unity at $t=0$);
the effective expansion rate, normalized by the Hubble constant of a
vacuum energy dominated universe.  We show the maximum,
minimum (omitted from the top panel), and volume-averaged value of the given quantities.
In the second subplot, in addition to
$\exp(\mathcal{N})$, we also show the cube-root of the spatial volume. 
We ignore apparent horizon interiors when calculating these quantities,
which accounts for the discontinuous features in the bottom two panels.
\label{fig:3d_ksq1_tseries}
}
\end{figure}

We choose $V(\phi)=\Lambda$ and initial data with scalar field perturbations of
a single wavenumber magnitude corresponding to $N=1$ in Eq.~(\ref{eq:phi_ic}),
and consider cases with different ratios of $k$ to initial effective Hubble
constant $H_0$.  As shown in Fig.~\ref{fig:rhoH_a}, for larger values of this
ratio ($k/H_0=4$ and 8), as the scale factor increases the average expansion
rate decreases like $\propto a^{-2}$, and the average gradient/kinetic energy of
the scalar field decreases as $\propto a^{-4}$ (where $a=e^{\mathcal{N}}$), as
would be expected in an FRW universe.  We find similar results for initial
configurations with more than one wavenumber ($N=5$ is also shown in
Fig.~\ref{fig:rhoH_a}). This continues until the kinetic/gradient energy is
subdominant, leading to a cosmological constant-dominated universe undergoing
exponential expansion.  As shown in the top of Fig.~\ref{fig:3d_ksq1_tseries},
though initially there is spatial variation in the energy density and rates of
expansion/contraction, these eventually go away. 

For smaller values of $k/H_0$ we are in the strong-field regime and there is
prompt gravitational collapse in some regions. In the cases with $k/H_0=1$ and 2
shown in the bottom of Fig.~\ref{fig:3d_ksq1_tseries}, the maximum energy
density rapidly increases and black holes form at $t\approx H_0^{-1}$ (we
emphasize that with the gauge choices made here, the time coordinate is
different from that of the usual FRW metric).  This is expected since the hoop
conjecture~\cite{Thorne72a} predicts that a trapped surface will form roughly
when the mass of an overdensity is comparable to the mass of a black hole of the
same size $\frac{4}{3}\pi G k^{-3}\rho \sim k^{-1}/2$, which is equivalent to $k
\sim H$. In terms of the amplitude of the scalar field fluctuations, this is 
also equivalent to $\delta \phi \sim M_P$.  
Because of the symmetry in the initial data and potential with respect
to positive and negative values of the scalar field, two identical apparent
horizons form in the periodic domain.  At the final time shown in
Fig.~\ref{fig:3d_ksq1_tseries}, the irreducible mass of each apparent horizons is
$\approx 2 k^{-1}$ and $\approx 0.6 k^{-1}$ for $k/H_0=1$ and 2, respectively.

We do not track the interior of the apparent horizons and avoid any issues with
singularities that may develop there by excising this region from the
computational domain.  However, ignoring the black hole interiors, even in these
cases there is still expansion on average and the volume-averaged density
decreases until the domain becomes vacuum energy dominated.
The proper distance between the black holes also increases roughly in
proportion to the average scale factor, indicating they are becoming more and
more isolated in an exponentially expanding universe (thus we end in a
situation similar to~\cite{Yoo:2014boa}). 

Similar results to those obtained with a cosmological constant potential should
hold in general for cases where the entire range of scalar field falls on flat
region of the potential.  We have explicitly verified that for
Eq.~(\ref{eq:notch_potential}) with $\lambda=1/\sqrt{6}$
~\cite{Kallosh:2013hoa}, and $\phi_0=6.2$ (corresponding to $\mathcal{N}\sim 60$
in homogeneous inflation), the results obtained are indistinguishable from those
presented above with $k/H_0=1$.  We next consider cases where the range of
scalar field values does not satisfy this condition, and hence the average value
of the scalar field $\phi_0$ will be important. For simplicity we will
concentrate on initial scalar field configurations with $N=1$ and $k/H_0=4$ and
vary $\phi_0$.   

\begin{figure}
\begin{center}
\vspace{-5pt}
\includegraphics[width=\columnwidth]{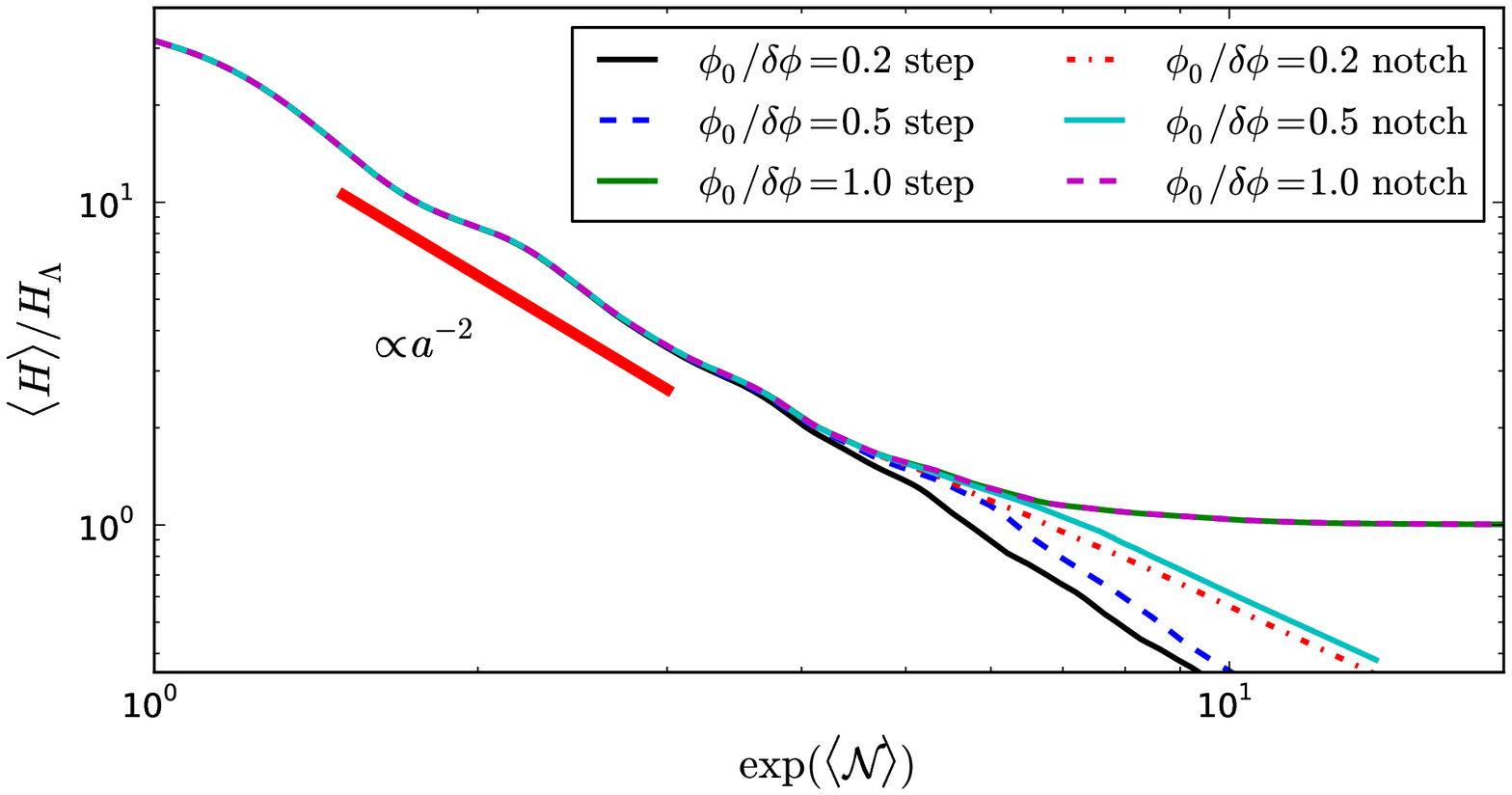}
\includegraphics[width=\columnwidth,draft=false]{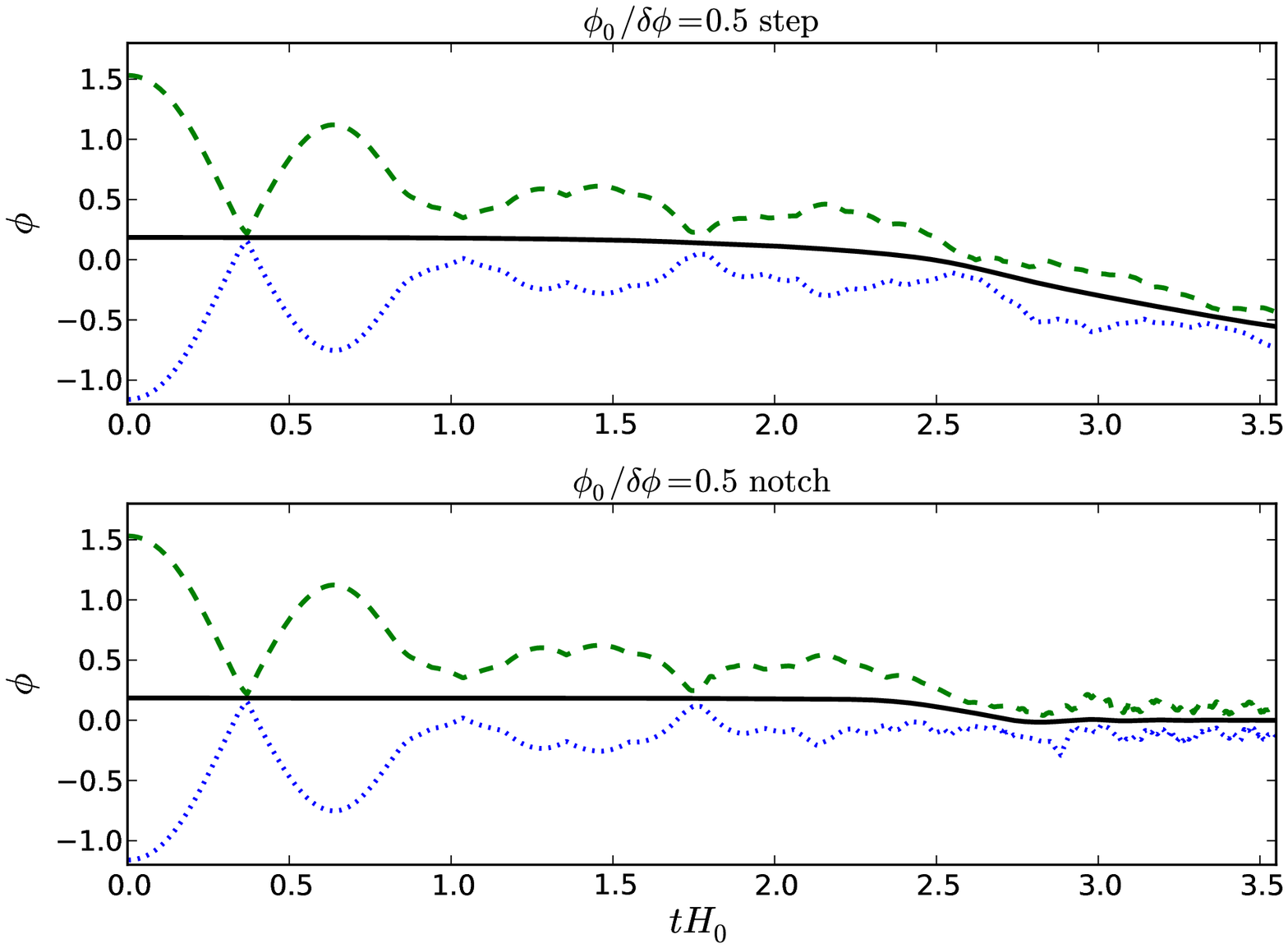}
\end{center}
\vspace{-5pt}
\caption{
The volume averaged expansion rate versus scale factor (top) and the minimum,
maximum, and volume-averaged values of the scalar as a function of time (bottom)
for cases with $k/H_0=4$, the step-like (Eq.~(\ref{eq:step_potential}) with
$\lambda = 200$) or notch-like (Eq.~(\ref{eq:notch_potential})
with $\lambda=100$) potentials, and various values of $\phi_0$. 
\label{fig:rhoH_a_step} 
\label{fig:rhoH_a_notch} }
\vspace{-5pt}
\end{figure}

In Fig.~\ref{fig:rhoH_a_step} we show several cases with the potential given by
Eqs.~(\ref{eq:step_potential}) with $\lambda = 200$, which gives a very good
approximation to a step function.  Of these, the one with the smallest average
scalar field value ($\phi_0=0.03675$) would give $\sim 60$ $e$-folds of
inflation before rolling down to a negative value in a homogeneous FRW model.
However in the inhomogeneous case we consider here, the value of the scalar
field everywhere becomes negative before the gradient energy of the scalar field
is negligible, and there is no phase of exponential expansion. This also happens
for the larger average value of $\phi_0/\delta \phi=0.5$. However, for
$\phi_0/\delta \phi=1$ the scalar field approaches a positive value
everywhere as the potential energy dominates and there is an inflationary
period.  

As also shown in Fig.~\ref{fig:rhoH_a_notch}, similar results are obtained using
the potential given by Eq.~(\ref{eq:notch_potential}) with $\lambda = 100$.  Here
again, for the cases where $\phi_0/\delta \phi$ is small, the scalar field moves
away from the inflationary plateau to a region of field space with lower
potential energy, in this case the narrow region at the minimum of the potential
at $\phi=0$. 

This can be seen analytically as follows.  In a homogeneous universe, time
derivatives of $\phi$ are proportional to the ``force" $V'(\phi)$.  When $\delta
\phi \neq 0$, spatially averaged time derivatives are proportional to the
averaged force $\la V' \ra$.  In the case of the step, when $\delta \phi >
\phi_0$, $\la V' \ra \sim V/\la \phi \ra$ and so  $\la \phi \ra$ is pushed
towards negative values parametrically faster than if $\delta \phi < \phi_0$.
In general, on a slow-roll plateau, $M_P V'/V \sim \sqrt{\epsilon} < 1$.   If
the fluctuations are large enough to reach the minimum of $V$, the spatially
averaged force is
$
\la V' (\phi) \ra \sim { V(\la \phi \ra) \over \la \phi\ra } = \( {M_P \over \la \phi \ra} \) \( { V'(\la \phi \ra)\over \sqrt{\epsilon}} \) > V'(\la \phi \ra)
$
unless $\la \phi \ra > M_P/\sqrt{\epsilon}$.  This is the origin of the
statement made in the introduction regarding  $\delta \phi < \Delta \phi$.  From
an Effective Field Theory (EFT) point of view, this result can be understood by
noticing that the effective potential for the homogeneous mode, which evolves
with a time scale of order Hubble, is obtained by integrating out the short (and
fast) modes which are classically excited. Due to the strong nonlinearities,
the effective potential can acquire a width of order the amplitude of the
classical modes.

One might expect that a potential with a sharp symmetric minimum could avoid
this tendency.  Consider the notch potential, \eqref{eq:notch_potential} with
$\lambda$ large.  Then $\la V' \ra$ is suppressed by $1/\lambda$, and so one
might expect a sharp notch to have only a small effect, which is indeed the case
in the absence of gravity.  However, with gravity included the combination of
Hubble friction and gravitational nonlinearities rapidly pulls $\phi$ into the
minimum. The rate at which $\la \phi \ra$ decreases is  almost independent of
$\lambda$ for large $\lambda$. From an EFT point of view, in an expansion in
$k/(aH)\gg 1$, we have that the leading effect vanishes, but higher orders do
not.

\section{Conclusions}
For a large class of examples, we find that exponential expansion occurs even
when the initial gradients are much larger than the potential energy.  The
possible exceptions are cases when the range of the initial inhomogeneous scalar
field values exceeds the inflationary plateau, in which case the outcome depends
on the details of the potential. This is illustrated by the potential with a
nearly flat  plateau for $\phi > 0$, and a relatively sharp ``step'' down to
zero at $\phi< 0$ given by \eqref{eq:step_potential} with $\lambda$ large: when
$\delta \phi \simgeq \phi_{0}$ the large $V'$ at the step near $\phi=0$ has a
strong effect on the time evolution of $\la \phi \ra$,  rapidly propelling it to
negative values and preventing inflation from beginning. We provide an
analytical understanding of this behavior.  We note, however, that this scenario
does not apply to generic examples of large-field inflation where $\phi_0$ is
large in units of $M_P$, but the fluctuations about this value are order one or
smaller. 

When the fluctuations in the inflaton field are contained within a flat region
of the potential, the potential will essentially act like a cosmological
constant, in which case in any region of the universe there are two possible
outcomes: recollapse after a finite time, or expansion until the vacuum energy
dominates and inflation begins.  With our initial conditions --- large
fluctuations of wavelength $\leq 2\pi/H_0$ in expanding universes that are flat
(or open) on average --- we find that both types of regions occur, with the
former leading to black holes, and the latter eventually dominating the volume.
Thus the amplitude of the initial-state fluctuations in $\phi$ is irrelevant
except in determining the time scales. {\it In particular, we find that in order
for inflation to start somewhere, there is no need to assume a Hubble-sized
homogeneous initial patch.} 

An important issue is how natural it might be to have an inflaton potential and
initial field range that satisfy the criteria laid out above. That question is
beyond the scope of this work and we do not address it here, though we refer the
reader
to~\cite{Starobinsky:1980te,Goncharov:1985yu,Bezrukov:2007ep,Carrasco:2015rva}
and references therein which discuss how cosmological attractor configurations
with the desired properties, such as approximate shift symmetric potentials, can
be constructed in supergravity, Higgs inflation, and other contexts.
Additionally, exploring a broader class of initial conditions, including
multiple fields, or the effects of homogeneous kinetic energy, will be
interesting follow-up work.

\acknowledgments

It is a pleasure to thank Tom Abel for initial collaboration in the project. The
work of MK is supported in part by the NSF through grant PHY-1214302, and he
acknowledges membership at the NYU-ECNU Joint Physics Research Institute in
Shanghai. The work by AL was supported by the SITP, by the NSF Grant PHY-1316699
and by the Templeton foundation grant Inflation, the Multiverse, and Holography.
LS is supported by DOE Early Career Award DE-FG02- 12ER41854 and by NSF grant
PHY-1068380.  Simulations were run on the Bullet Cluster at SLAC and the Orbital
Cluster at Princeton University.

\appendix
\section{Details of numerical methods}
\label{numerical}
\begin{figure}[t!]
\begin{center}
\includegraphics[width=\columnwidth,draft=false]{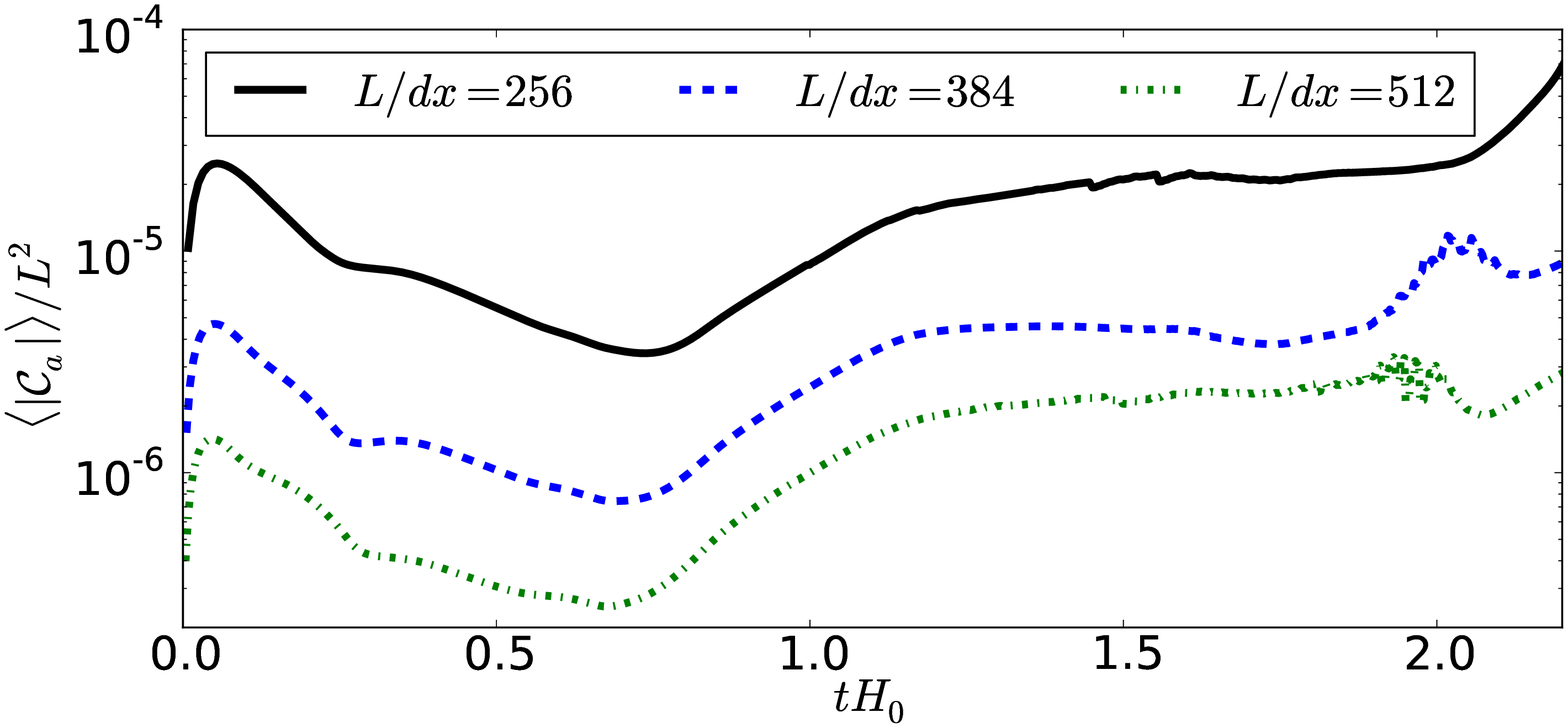}
\end{center}
\caption{
A resolution study of the $k/H_0=1$ ($N=1$) case.  The scaling of the norm of
the violation of the generalized harmonic constraint ($C_a=H_a-\Box x_a$ where
$H_a$ are the source functions) with resolution is consistent with between third
and fourth order convergence.
\label{fig:cnst_conv} }
\end{figure}

\begin{figure}[t!]
\begin{center}
\includegraphics[width=\columnwidth,draft=false]{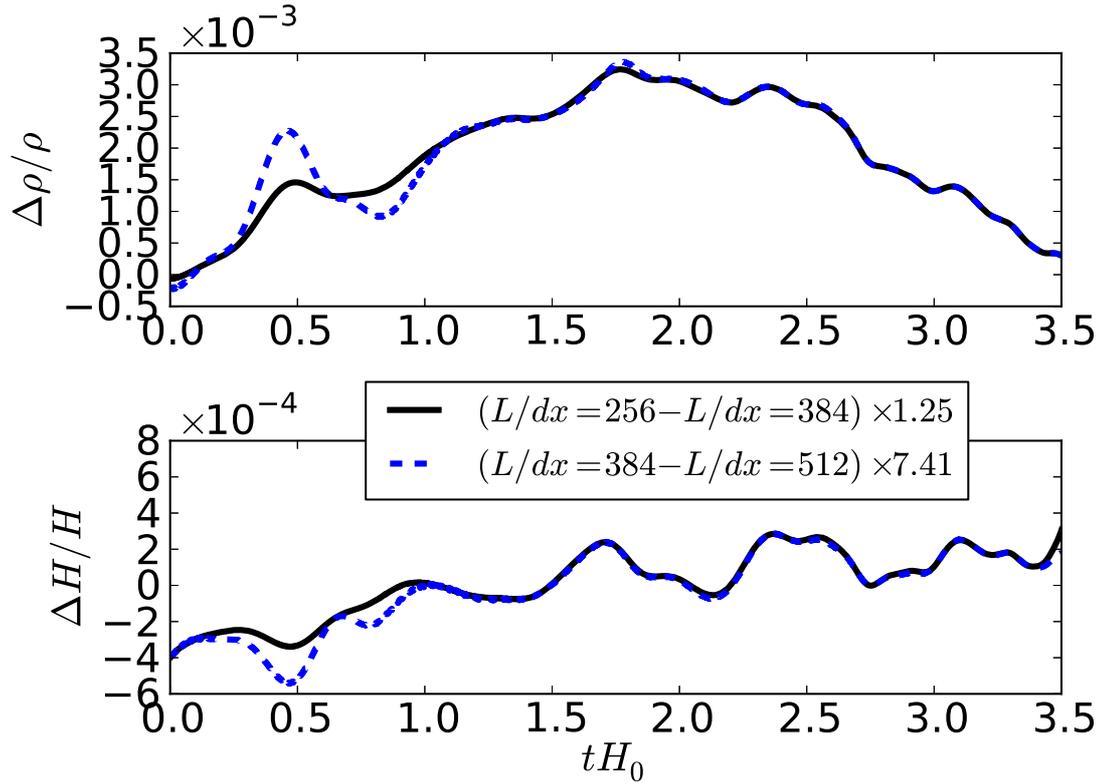}
\end{center}
\caption{
A resolution study of the $k/H_0=4$ ($N=1$) case.  
Shown is the relative difference in the volume-averaged energy density (top) and
expansion rate (bottom) between different resolutions. The values are scaled
to show the relative error in the $L/dx=256$ case, assuming fourth-order
convergence.
\label{fig:error} }
\end{figure}

In order to numerically simulate an inhomogeneous cosmology, we solve the Einstein field
equations coupled to the inflaton $\phi$ in a periodic domain.  The
field equations are evolved in the generalized harmonic formulation, using the code
described in~\cite{Pretorius:2004jg, code_paper}.  We use the same variation of
the damped harmonic gauge~\cite{Choptuik:2009ww,Lindblom:2009tu} as
in~\cite{East:2012mb}.  The inflaton equation of motion $\Box \phi = V' $
(where $V$ is the potential) is evolved using the same fourth-order finite
difference stencils and Runge-Kutta time stepping as the metric.  As the
simulations evolve and the metric components grow due to expansion, we
dynamically adjust the timestep size in proportion to the decreasing global
minimum of $\gamma^{1/6}/\alpha$ (where $\gamma$ is the determinant of the spatial metric
and $\alpha$ is the lapse) in order to avoid
violating the Courant-Friedrichs-Lewy condition.  The simulations are performed
with between 256 and 512 points across each linear dimension to establish
numerical convergence and estimate error.  The convergence of the constraints is
shown for the most strong-field case considered above in
Fig.~\ref{fig:cnst_conv}. In Fig.~\ref{fig:error}, we show the truncation error
in the volume-averaged energy density and expansion rate for the $k/H_0=4$ case,
which are sub-percent level even at the lowest resolution.


\bibliographystyle{JHEP}
\bibliography{ref}

\end{document}